\documentclass[sigconf]{acmart}
\usepackage{xcolor}
\usepackage{listings}
\usepackage[export]{adjustbox}
\usepackage{graphicx}
\usepackage{pifont}
\newcommand{\cmark}{\ding{51}} 
\newcommand{\xmark}{\ding{55}} 
\usepackage{color}
\usepackage{comment}
\usepackage{booktabs}  
\usepackage{tabularx}  
\usepackage{array}     
\renewcommand{\arraystretch}{1.3} 
\definecolor{mygreen}{rgb}{0,0.6,0}
\definecolor{mygray}{rgb}{0.5,0.5,0.5}
\definecolor{mymauve}{rgb}{0.58,0,0.82}
\usepackage{listings}
\AtBeginDocument{%
  }

\setcopyright{acmlicensed}
\copyrightyear{2018}
\acmYear{2018}
\acmDOI{XXXXXXX.XXXXXXX}
\acmISBN{978-1-4503-XXXX-X/2018/06}

\extrafloats{500}
\begin{document}


\title{Integrating LLMs in Software Engineering Education: Motivators, Demotivators, and a Roadmap Towards a Framework for Finnish Higher Education Institutes}



\author{Maryam Khan}
\affiliation{%
  \institution{Software Engineering Department,
Lappeenranta-Lahti University of
Technology}
  \city{Lappeenranta}
  \country{Finland}}
\email{maryam.khan@student.lut.fi}

\author{Muhammad Azeem Akbar}
\affiliation{%
  \institution{Software Engineering Department,
Lappeenranta-Lahti University of
Technology}
  \city{Lappeenranta}
  \country{Finland}}
\email{azeem.akbar@lut.fi}

\author{Jussi Kasurinen}
\affiliation{%
  \institution{Software Engineering Department,
Lappeenranta-Lahti University of
Technology}
  \city{Lappeenranta}
  \country{Finland}}
\email{Jussi.Kasurinen@lut.fi}

\renewcommand{\shortauthors}{Khan et al.}

\begin{abstract}
  The increasing adoption of Large Language Models (LLMs) in software engineering education presents both opportunities and challenges. While LLMs offer benefits such as enhanced learning experiences, automated assessments, and personalized tutoring, their integration also raises concerns about academic integrity, student over-reliance, and ethical considerations. In this study, we conducted a preliminary literature review to identify motivators and demotivators for using LLMs in software engineering education. We applied a thematic mapping process to categorize and structure these factors (motivators and demotivators), offering a comprehensive view of their impact. In total, we identified 25 motivators and 30 demotivators, which are further organized into four high-level themes. This mapping provides a structured framework for understanding the factors that influence the integration of LLMs in software engineering education, both positively and negatively. As part of a larger research project, this study serves as a feasibility assessment, laying the groundwork for future systematic literature review and empirical studies. Ultimately, this project aims to develop a framework to assist Finnish higher education institutions in effectively integrating LLMs into software engineering education while addressing potential risks and challenges.
\end{abstract}




\maketitle
\section{INTRODUCTION}
The rise of artificial intelligence (AI) has transformed various industries towards automation \cite{acemoglu2018artificial}, including healthcare \cite{shaheen2021applications}, finance \cite{cao2022ai}, entertainment \cite{forbus2002guest}, and notably education \cite{zhai2021review,jin2025generative}, the primary focus of this study. LLMs lead the AI transformation and revolution, based on their core characteristics such as generating human-like responses to various prompts. LLMs are defined using deep learning algorithms and Natural Language Processing (NLP) techniques to develop coherent and relevant responses \cite{akbar2023ethical}. The models are based on transformer architectures and pre-trained on large amounts of textual data.

The unique characteristics and expanding applications of LLMs introduce a spectrum of potential opportunities and distinct challenges, particularly in the landscape of software engineering education \cite{he2024llm}. LLMs hold significant potential for use in academia, especially in software engineering education—performing various software engineering activities. However, software engineering education is at a crucial stage due to AI-based technologies such as LLMs, which find applications in chatbots, content creation, and code development tools \cite{kirova2024software}. Therefore, the increasing use of LLMs necessitates restructuring software engineering education to integrate AI technologies \cite{kirova2024software, fan2023large}. Modern software engineering education programs need to holistically incorporate the unique applications of LLMs, ensuring that students are educated about the challenges and opportunities they present \cite{daun2023chatgpt}.

Various research studies have been conducted focusing on the integration of AI technologies in education, such as software engineering and computer science \cite{banerjee2025understanding, feldt2018ways, lo2023impact}. However, the rapidly evolving AI landscape presents unique challenges for designing curricula and pedagogy. Incorporating AI tools, machine learning techniques, and the ethical implications of LLMs into software engineering education are active areas of research discussion and still remain relatively unexplored.

As highlighted in recent research, Finland progressive education system, known for its innovation and inclusivity, provides an ideal context for developing a tailored framework that aligns with national pedagogical goals and regulatory requirements \cite{fattahibavandpour2024advancing}. Recent work also emphasizes the importance of usability-focused AI tools in education, particularly in empowering teachers with AI-driven course management and curriculum planning solutions. These tools not only support educators but also enhance sustainable and work-life-relevant education by integrating LLM capabilities into existing learning platforms \cite{macias2024empowering}. However, despite these advancements, the application of LLMs in software engineering education in Finland remains highly unexplored. While AI-driven tools have been adopted in digital humanities and curriculum planning, their structured integration into SE education, particularly in Finnish universities, lacks systematic research and implementation.

It is acknowledged that while LLMs are powerful tools, they are not without limitations. Concerns have been raised about their susceptibility for biased decision-making, ethical issues, and their "black box" nature \cite{bender2021dangers}. These challenges are not just academic concerns; their practical implications make them highly relevant for software engineering students. The growing discussion on AI ethics has also captured significant attention in academia. This is reflected in the literature, where numerous studies have been conducted to emphasize the importance of understanding AI principles such as transparency, privacy, accountability, fairness, and autonomy \cite{khan2023ai}. These ethical principles are crucial not only to the AI field but also to software engineering students who are primarily involved in developing AI-powered software solutions. It is essential to integrate the ethical aspects of LLMs into the software engineering curriculum \cite{khan2022ethics}.

Therefore, this work outlines how software engineering institutes in Finland are incorporating LLMs into their educational systems to attain the promising benefits of AI technologies. This integration reflects the need for a deep understanding of these technologies—LLMs. This research focuses on several key areas of knowledge to provide a framework for Finnish software engineering research institutes. These include the challenges associated with integrating LLMs into educational systems, the causes of these challenges, and the best solutions. Furthermore, we aim to understand the existing guidelines and frameworks developed for integrating LLMs in education across other domains. Based on this, we will define a framework specific to software engineering education for Finnish higher education institutes. In the long term, this work aims to contribute to the ongoing discussions and dialogues concerning the future of software engineering education in an AI environment and serve as a set of guidelines for educational institutes, policymakers, students, and curriculum developers. To achieve the study aims and objectives, we have developed the core research questions and their rationale given in Table \ref{table:tab1}.

In this article, we present the preliminary literature review of this project to address RQ1.1, focusing on discussions surrounding the use of LLMs in software engineering education to establish the feasibility of this project. We identified 25 motivating and 30 demotivating factors influencing the adoption of LLMs in software engineering education. Furthermore, a thematic mapping of these factors was conducted to categorize them into high-level themes. In the future, we aim to conduct a comprehensive systematic literature review (SLR) and empirical studies to develop the final framework (See Section \ref{Sec:Research_Methodology}).

The rest of the paper is structured as follows: The research methodology of the project and the preliminary review protocol are briefly discussed in Section \ref{Sec:Research_Methodology}. The study results are presented in Section \ref{Sec:Results}. The implications of this study are detailed in Section \ref{Sec:Implications}, followed by a discussion on validity threats in Section \ref{Sec:Threat_Validity)}. Finally, we conclude the study and outline future research directions in Section \ref{sec:Conlusions}.

\begin{table*}[t]  
\caption{Research Questions of this Project}
\label{table:tab1}
\centering
\small  
\renewcommand{\arraystretch}{1.2}  

\begin{tabular}{|m{1.2cm}|m{5cm}|m{8cm}|}  
\hline
\textbf{RQ\#} & \textbf{Research Question} & \textbf{Rationale} \\
\hline
RQ1 & How can LLMs be effectively integrated into Finnish higher education institutions to enhance teaching and learning in software engineering education? & 
This research question aims to explore the integration of LLMs into Finnish higher education to potentially transform and enhance the pedagogical practices in software engineering education. It aims to identify effective strategies and practices for deploying LLMs that could improve both teaching methods and student learning outcomes. The focus on Finnish institutions provides context-specific insights that may lead to tailored solutions in educational technology. \\
\hline
RQ1.1 & What are the motivators and demotivators of integrating LLMs into software engineering education? & 
This research sub-question investigates the specific success factors (motivators) and challenges (demotivators) that should be considered when using LLMs in software engineering educational programs in Finnish higher education institutes. It aims to uncover technical, ethical, and pedagogical motivators and demotivators in using AI-driven tools within a curriculum traditionally focused on highly technical and specialized skills. Understanding these factors is crucial for developing strategies to effectively implement LLMs and increase their educational benefits. \\
\hline
RQ1.2 & What are the causes of the identified challenges and mitigation practices? & 
This question is defined to investigate the root causes of the challenges identified in RQ1.1. It seeks to understand the underlying factors contributing to these challenges and to explore potential mitigation practices. It will help in developing effective strategies and solutions that will ensure a successful integration of LLMs into educational frameworks (RQ3).  \\
\hline
RQ2 & What are the existing guidelines/frameworks in other domains that could assist higher education institutions in implementing LLMs? & 
This RQ aims to explore existing guidelines and frameworks in various domains (such as computer science, business, etc.) and contexts (other countries) that could inform and support the integration of LLMs into software engineering higher education settings in Finland. \\
\hline
RQ3 & What would be a framework to assist Finnish higher education institutions in implementing LLMs? & 
The final RQ aims to develop a specific framework to assist Finnish higher education institutions in effectively integrating LLMs into software engineering education. It will provide a structured approach that addresses the unique educational, cultural, and technical contexts of Finnish institutions. By establishing a clear framework, the question seeks to ensure that the deployment of LLMs enhances teaching and learning outcomes while also navigating any potential challenges specific to the local academic environment. The inputs of this research question are RQ1-RQ2. \\
\hline
\end{tabular}
\end{table*}

\section{Research Methodology} \label{Sec:Research_Methodology}
To address the study research questions, we will use mixed research methods, including a preliminary review (traditional literature survey) in the first phase of the research (this study), Systematic Literature Review (SLR), and empirical studies (questionnaire survey, interviews, and case study) in the second and final phase. The core steps of the proposed research methodology are discussed as follows and presented in Figure \ref{fig:figure1}:

\begin{figure*}
    \centering
    \includegraphics[width=0.5\textwidth]{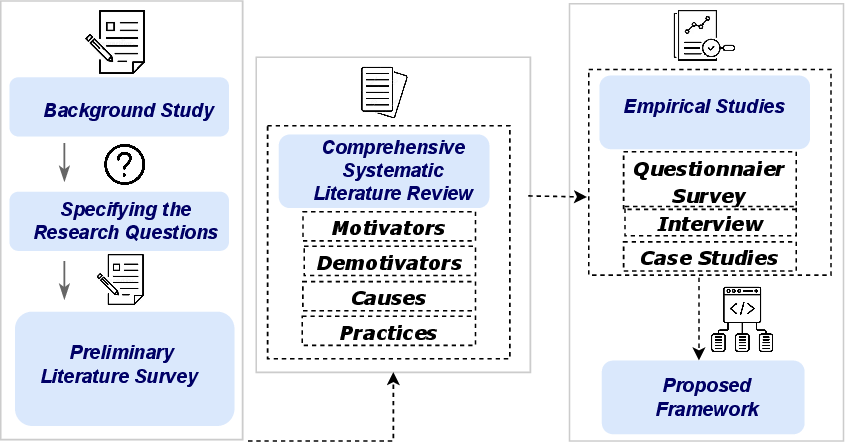}
    \caption{Proposed Research Process}
    \label{fig:figure1}
\end{figure*}

\begin{itemize}
\item \textbf{Preliminary Literature Survey}  
In this article, we conducted a preliminary literature survey using the traditional literature survey approach (see Section \ref{Traditional LR}) to identify motivators and demotivators of using LLMs in software engineering education to answer (RQ1.1). This preliminary study is conducted to evaluate the feasibility of this project in the long term. Additionally, it helps in identifying gaps in existing research and refining our research objectives for more targeted investigations. The insights gained from this phase provide a foundational understanding that informs subsequent stages of the study.

   \item \textbf{Systematic Literature Review (SLR)}  
    The SLR approach is primarily chosen to explore the existing literature and provide a comprehensive overview of motivators, demotivators, their causes, and mitigation practices of using LLMs in Finnish software engineering institutes (RQ1.1, RQ1.2). . This phase builds upon the insights gained from the Preliminary Literature Survey, ensuring a structured and in-depth investigation of relevant studies. Following the principles of evidence-based software engineering, we ensure a rigorous review process by systematically examining all primary studies related to the research questions in an unbiased and iterative way \cite{kitchenham2007guidelines}.

    \item \textbf{Empirical Studies}  
    The SLR findings will be empirically evaluated by conducting a questionnaire-based survey, complemented with interviews and a case study:
    \begin{itemize}
        \item \textbf{Questionnaire Survey:} A detailed survey \cite{punter2003conducting}, incorporating both closed and open-ended questions, will be conducted with teachers, students, and IT administrators to evaluate the SLR findings, including the identified motivators, demotivators, their thematic mapping, causes of demotivators, and best practices. Additionally, the survey will help identify new motivators, demotivators, causes, and best practices that are not discussed in the existing literature, further enriching the SLR findings. Moreover, we will validate the thematic mapping of motivators and demotivators based on the perceptions of survey participants. Finally, the descriptive analysis will be used to analyze the collected data \cite{kitchenham2002preliminary}.

        \item \textbf{Interviews:} Comprehensive interviews with experts in software engineering education will be conducted to validate the questionnaire survey findings and understand their perspectives about the findings of previous phases. To analyze the interview data, the Grounded Theory (GT) \cite{glaser2017discovery} approach will be used to define and derive the core codes from raw data and develop their mapping.

        \item \textbf{Case Studies:} Based on the interview findings, the framework will be finalized and its real-world effectiveness evaluated using a case study approach \cite{runeson2009guidelines}. Several Finnish universities will be targeted to deploy the framework and assess its effectiveness. Furthermore, it will be comparatively analyzed to determine if there are any existing practices of using LLMs in software engineering education. Eventually, the findings from Finland will be compared with those from other countries that have adopted similar technologies in education.
    \end{itemize}
\end{itemize}

In this study, we explicitly focus on the first phase of the research by conducting a preliminary traditional review (see Section \ref{Traditional LR}) to evaluate the feasibility of the project by answering RQ1.1. 

\section{Proposed Research Approach} \label{Traditional LR}
In the long term, we plan to conduct an SLR,  empirical and case studies as described in Section 2. However, to assess the significance of the proposed project, we conducted a preliminary traditional literature review to understand the key motivators and demotivators (RQ1.1) of using LLMs in software engineering education. The aim of this study is to investigate the factors—both motivators and demotivators—that positively and negatively influence the integration of LLMs in software engineering education. In this literature survey, the snowballing approach is adopted, which consists of backward snowballing (selecting studies based on their references) and forward snowballing (identifying studies that cite the selected studies) \cite{wohlin2014guidelines}. This approach gradually increases the pool of relevant studies. The studies identified through snowballing focus on the motivating factors (success factors) and demotivating factors (challenges) associated with using LLMs in software engineering education. We also included studies that do not explicitly mention these factors but focus on utilizing LLMs in software engineering education \cite{niazi2015systematic, khan2021agile}. However, extracting these factors from such studies is relatively challenging, requiring an extensive and detailed review \cite{niazi2015systematic}.

To ensure a broad search across digital repositories, we used Google Scholar as the primary search engine. It provides integrated access to various information technology and education-related repositories such as IEEE Xplore, Springer, ACM Digital Library, Wiley, etc. \cite{khan2021agile}. This approach ensures wide coverage and reduces the risk of missing relevant studies.

The first and second authors (supervisor) conducted the initial review process and resolved any disagreements in study selection through mutual discussion and consensus. Ultimately, a total of 45 studies were selected using the snowballing approach. These studies were preliminarily analyzed to extract a list of motivators and demotivators. 

The first and second authors reviewed all the selected studies and developed a list of motivators and demotivators for using LLMs in software engineering education. To eliminate interpersonal biases in the review process conducted by the authors, two external reviewers were invited. The reviewers were asked to randomly select ten studies from the list of selected studies and perform the review following the same procedure as the authors.

Finally, Kendall’s coefficient of concordance (W) test is performed to assess interpersonal biases between the authors and the external reviewers. Kendall’s coefficient of concordance (W) is a well-known statistical test used to evaluate the degree of agreement among a group of people assessing a set consisting of n variables \cite{sidney1957nonparametric}. The W score ranges from 0 to 1, where 0 indicates complete dis-agreement and 1 represents strong agreement. We used R-3.6.3 to calculate the W value, and the results, shown in Table \ref{tab:kendall_w} (W = 0.82), indicate that the degree of agreement between the authors and the invited reviewers in the review process is positive. This suggests that there were no interpersonal biases in the review process performed by the authors. The R code developed to calculate the W value is as follows:

\begin{lstlisting}[language=R, basicstyle=\ttfamily\footnotesize, frame=single, captionpos=b, breaklines=true, breakatwhitespace=true, columns=flexible]
library(DescTools)
# Define the rankings given by two external reviewers and authors (10 studies)
LLMs_SE_Education <- data.frame(
  external_rev1 = c(6, 5, 7, 1, 10, 4, 8, 3, 9, 2),
  external_rev2 = c(6, 7, 5, 2, 1, 4, 10, 9, 8, 3),
  authors       = c(4, 5, 7, 1, 2, 3, 15, 9, 8, 6)
)
# Calculate Kendall's coefficient of concordance (W)
result <- KendallW(LLMs_SE_Education, correct=TRUE, test=TRUE)
# Print results
print(result)
\end{lstlisting}

\begin{table}[H]  
\caption{Kendall’s Coefficient of Concordance (W) Test Results for LLMs in Software Engineering Education}
\label{tab:kendall_w}
\centering
\small  
\renewcommand{\arraystretch}{1.2}  

\resizebox{\columnwidth}{!}{  
\begin{tabular}{|l|c|c|c|c|c|c|}
\hline
\textbf{Data Set} & \textbf{Chi-Square} & \textbf{df} & \textbf{Vars} & \textbf{Eval} & \textbf{P-val} & \textbf{W} \\
\hline
LLMs in SE Education & 34.733 & 14 & 15 & 3 & 0.001 & 0.826 \\
\hline
\end{tabular}
}
\end{table}

We included the list of all 45 selected studies in the replication package \cite{khan2025motivators} prepared for this study. It is not provided in the article due to page limitations.
\section{Results} \label{Sec:Results}
\subsection{Identified Motivators}
Following the review protocol discussed in Section 3, we identified a total of 25 motivators that could positively affect the use of LLM tools in software engineering education. To present these motivators in a meaningful structure, we conducted a thematic analysis following the guidelines defined by Braun and Clarke \cite{braun2006using}. Thematic analysis is a well-known approach in social sciences for analyzing data to identify and report patterns (themes). The process consists of five core phases: (1) familiarization with the data (noting patterns, meanings, and initial ideas to develop a deep understanding of the content), (2) generating initial codes (systematically coding the data by identifying features of interest across the dataset), (3) searching for themes (sorting codes into potential themes by identifying broader patterns), (4) reviewing themes (checking the coherence of themes concerning the coded extracts), and (5) defining themes (identifying the essence of each theme and refining its scope).

\begin{figure*}[t]  
    \centering
    \includegraphics[width=0.7\textwidth]{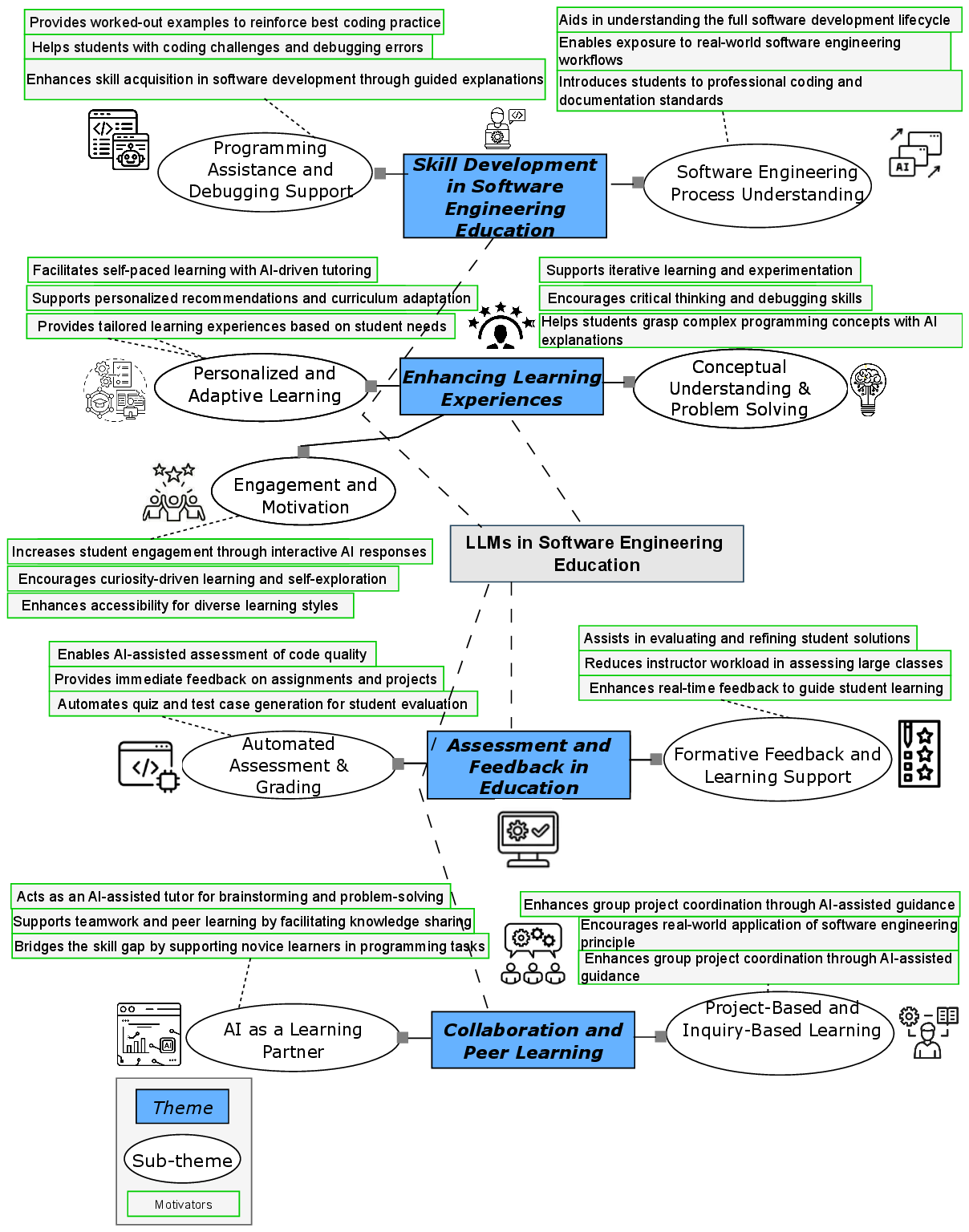}  
    \caption{Thematic Mapping of Identified Motivators}
    \label{fig:figure2}
\end{figure*}

In this study, the first author primarily led the thematic mapping process and defined the themes and sub-themes of the identified motivators and demotivators. The second author continuously reviewed the mapping activities and collaborated with the first author to resolve any differences or address potential issues such as bias. Finally, the identified motivators are mapped across 9 sub-themes and 4 high-level themes as shown in Figure \ref{fig:figure2}. 

We now summarize the high-level themes and the details are as follows:

\subsubsection{Enhancing Learning Experiences}
Enhancing Learning Experiences emerged as an important high-level theme consisting of total 9 motivating factors and 3 sub-themes. It mainly focuses on how LLMs contribute to creating a more personalized, interactive, and engaging educational environment for software engineering students \cite{cowan2023enhancing, jovst2024impact}. The use of  LLMs in software engineering education significantly enhances learning experiences by fostering personalized and adaptive learning, engagement, and conceptual understanding \cite{cowan2023enhancing}. LLMs provide tailored learning experiences, enabling students to learn at their own pace with AI-driven tutoring and adaptive recommendations. This level of customization ensures better knowledge retention and curriculum alignment to individual student needs \cite{korpimies2024unrestricted}. Additionally, LLMs increase student engagement and motivation by facilitating interactive learning environments that encourage curiosity-driven exploration, making learning more dynamic and accessible to diverse learning styles \cite{cambaz2024use}. The ability of LLMs to simplify complex programming concepts with clear explanations fosters conceptual understanding and problem-solving skills, helping students develop critical debugging abilities and embrace iterative learning through experimentation \cite{cowan2023enhancing, jovst2024impact}. These advantages make AI a valuable tool in modern education, bridging the gap between students’ learning needs and the evolving demands of software engineering.

\subsubsection{Assessment and Feedback in Education}
LLMs play a pivotal role in improving assessment and feedback mechanisms by automating grading, providing real-time formative feedback, and reducing instructor workload \cite{meissner2024evalquiz}. AI-powered automated assessment and grading systems offer instant feedback on assignments and projects, enabling students to quickly identify mistakes and improve their work \cite{rasnayaka2024empirical, lyu2024evaluating}. Moreover, AI-assisted test case generation and evaluation contribute to more efficient and scalable assessment processes, particularly for large classes. Beyond grading, LLMs support formative feedback and learning guidance, allowing students to refine their solutions through real-time, interactive feedback \cite{meissner2024evalquiz}. This not only enhances learning but also alleviates the burden on educators by automating the evaluation of student progress. The ability of AI to provide consistent and objective feedback ensures a fair assessment process, making AI-assisted evaluation an essential tool for modern education \cite{meissner2024evalquiz}. Additionally, AI-enhanced assessments can detect patterns in student performance, helping instructors tailor their teaching strategies to address common difficulties \cite{hanifi2023chatgpt}. This data-driven approach ensures that students receive targeted support for their weaker areas, leading to more effective knowledge acquisition. Furthermore, LLMs can simulate real-world evaluation scenarios, preparing students for professional coding environments where automated testing and debugging are standard practices. By integrating AI into assessment strategies, educators can foster a more efficient, data-driven, and student-centered evaluation system that aligns with industry expectations.  

   \subsubsection{Collaboration and Peer Learning} 
   LLMs contribute to collaboration and peer learning by acting as AI-assisted learning partners and enhancing project-based, inquiry-based learning approaches \cite{cambaz2024use, vierhauser2024towards}. AI serves as a virtual tutor for brainstorming and problem-solving, allowing students to bridge skill gaps and gain structured guidance in programming tasks \cite{wang2024enhancing}. This fosters an environment where teamwork and knowledge sharing become more effective, as AI can facilitate discussions, provide relevant insights, and help students understand complex topics collaboratively. Additionally, AI improves group project coordination, offering structured guidance that ensures real-world application of engineering principles while supporting agile and iterative methodologies. This integration of AI in teamwork-driven learning encourages students to engage more actively in peer learning, strengthening both technical and collaborative skills essential for software engineering professionals \cite{kharrufa2024llms}. Moreover, AI-driven learning environments can enable students to work asynchronously on projects, improving flexibility and accessibility for remote or international teams. AI can also provide dynamic feedback on collaborative efforts, helping teams refine their work based on AI-generated insights and best practices \cite{marques2024using}. By leveraging AI as a knowledge-sharing tool, students can expand their perspectives, learning from globalized AI resources beyond their immediate academic environment \cite{santos2024generative}. This creates a more interconnected and informed student community, fostering a culture of collective learning and innovation in software engineering education. 
    
    \subsubsection{Skill Development in Software Engineering Education}
    LLMs significantly enhance skill development in software engineering education by offering programming assistance, debugging support, and software engineering process understanding. AI helps students tackle coding challenges by providing contextual explanations and step-by-step debugging guidance, making complex software development tasks more manageable \cite{meissner2024evalquiz, herden2024integration}. By offering worked-out examples and best coding practices, AI ensures that students develop robust programming habits and reinforce essential software development skills \cite{boguslawski2025programming}. Furthermore, AI supports a comprehensive understanding of the software engineering lifecycle, guiding students through documentation standards, software development processes, and real-world engineering workflows \cite{pereira2024leveraging}. Exposure to industry-relevant software engineering methodologies through AI-driven insights prepares students for practical applications of their learning, ensuring they are equipped with the knowledge and skills necessary for professional success. Additionally, AI tools can simulate industry coding environments, allowing students to practice within real-world constraints such as efficiency, scalability, and security considerations \cite{zhang2023survey}. This hands-on experience with AI guidance helps bridge the gap between theoretical knowledge and applied engineering skills, making students better prepared for the demands of the workforce. Moreover, AI-driven debugging assistance accelerates the learning curve for novice programmers by highlighting common mistakes and suggesting optimized solutions \cite{cowan2023enhancing, cambaz2024use, wang2024enhancing}. By integrating AI into skill development strategies, software engineering education can produce more competent and industry-ready graduates, equipping them with the problem-solving capabilities essential for future technological advancements.

    In summary, while LLMs significantly enhance learning experiences, assessment processes, collaboration, and skill development, their effectiveness depends on structured implementation strategies that prevent over-reliance on AI-generated solutions. Educators must design curricula that leverage AI as a complement to traditional teaching methods rather than as a replacement. Ethical concerns regarding AI-assisted assessments, academic integrity, and student dependence on automation must also be addressed. Ensuring that LLMs serve as facilitators of deeper understanding rather than shortcuts to learning outcomes is crucial. The future of software engineering education will benefit from AI-driven tools, but their role should be carefully managed to maintain a balance between technological support and the cultivation of essential problem-solving skills among students.

\subsection{Identified Demotivators}
\begin{figure*}[t]  
    \centering
    \includegraphics[width=0.7\textwidth]{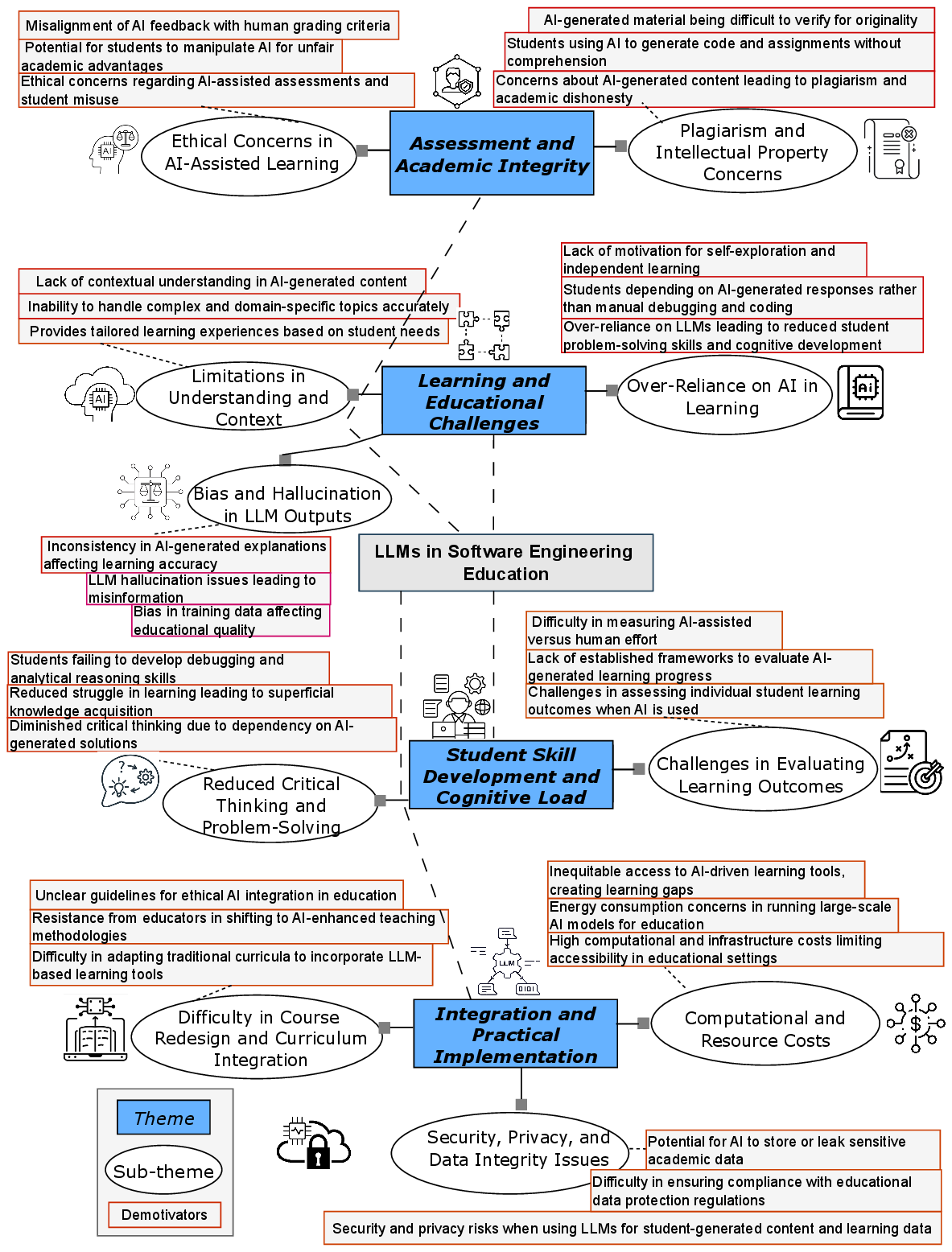}  
    \caption{Thematic Mapping of Identified Demotivators}
    \label{fig:figure3}
\end{figure*}

Following the defined review protocol (Section \ref{Traditional LR}), we identified 30 demotivators that could hinder the integration of LLMs in software engineering education. Additionally, we performed a thematic mapping of these demotivators, as outlined in Section 4.1. Our analysis resulted in the development of four high-level themes and ten sub-themes of demotivating factors, as illustrated in Figure \ref{fig:figure3}. Below, we discuss each high-level theme in detail.

 \subsubsection{Learning and Educational Challenges}

    The high-level theme of Learning and Educational Challenges highlights critical concerns regarding the integration of LLMs in software engineering education, categorized into three key sub-themes: limitations in understanding and context, bias and hallucination in LLM outputs, and over-reliance on AI in learning. LLMs struggle with contextual understanding, leading to inaccurate handling of domain-specific topics and difficulties in maintaining conversation coherence, which can hinder deep learning and effective knowledge transfer \cite{sallou2024breaking}. Furthermore, bias and hallucination in LLM outputs pose a significant risk, as the presence of biased training data and misinformation can affect the credibility and quality of educational content, potentially misleading students \cite{jalil2023chatgpt, hussein2024usability}. Inconsistencies in AI-generated explanations further exacerbate these issues, impacting learning accuracy and reliability. The third challenge, over-reliance on AI in learning, raises concerns about students becoming overly dependent on AI-generated responses rather than engaging in critical thinking, debugging, and problem-solving \cite{ozkaya2023application}. This reliance can lead to a decline in cognitive development and self-directed learning, reducing students' motivation for independent exploration \cite{cambaz2024use, santos2024generative}. Collectively, these factors emphasize the need for a balanced approach in integrating LLMs into software engineering education, ensuring that AI complements rather than replaces essential learning and problem-solving skills.

    \subsubsection{Assessment and Academic Integrity}

    The integration of LLMs in software engineering education raises significant concerns regarding assessment and academic integrity, particularly in plagiarism, intellectual property, and ethical concerns in AI-assisted learning. AI-generated content can lead to increased instances of academic dishonesty, as students may use LLMs to complete assignments without genuine comprehension, making it difficult to verify originality \cite{hussein2024usability}. This threatens the value of assessments and undermines learning objectives. Additionally, ethical concerns arise in AI-assisted assessments, where AI feedback may misalign with traditional grading criteria, leading to potential misuse by students to gain unfair advantages \cite{samhan2024review}. The ease of manipulating AI-generated solutions also creates risks in maintaining fair evaluation processes \cite{gao2024fairness}. The growing dependence on AI for content generation further blurs the line between original student work and AI-assisted outputs, making it imperative for institutions to develop stronger detection mechanisms for AI-generated plagiarism. Moreover, the risk of students misusing AI to bypass academic effort threatens the development of genuine expertise in software engineering disciplines \cite{bender2021dangers}. To mitigate these risks, universities must implement strict AI policies, promote academic integrity awareness, and redesign assessments to encourage critical thinking over rote AI-generated responses.

    \subsection{Student Skill Development and Cognitive Load}

    A key challenge in adopting LLMs in education is their impact on student skill development and cognitive load, especially in critical thinking, problem-solving, and evaluating learning outcomes \cite{wang2025effects}. Dependence on AI-generated solutions may diminish students ability to engage in critical reasoning and debugging, leading to superficial knowledge acquisition rather than deep learning. The automation of learning tasks reduces cognitive effort, which, while lowering barriers for understanding, may also stunt problem-solving development \cite{toma2024effects}. Furthermore, assessing individual learning progress becomes difficult, as distinguishing between AI-assisted knowledge and independently acquired skills remains a challenge. The lack of established frameworks for evaluating AI-based learning outcomes further complicates assessment strategies, raising concerns about the effectiveness of AI-driven education in fostering essential analytical and problem-solving skills. Without proper guidance on the responsible use of AI, students may rely on LLMs to shortcut difficult tasks \cite{joshi2023great}, diminishing their resilience in tackling complex coding and debugging challenges. Additionally, the ease of access to AI-generated answers may reduce intellectual curiosity, preventing students from exploring alternative problem-solving strategies. A well-structured integration of AI should focus on enhancing learning rather than replacing cognitive engagement, ensuring that AI acts as a supportive tool rather than a substitute for fundamental engineering competencies.

    \subsubsection{Integration and Practical Implementation}

    The practical implementation of LLMs in software engineering education is hindered by challenges in course redesign, computational costs, and security concerns. Adapting existing curricula to incorporate LLM-based learning tools is complex and resource-intensive, as educators often resist shifting to AI-driven methodologies due to a lack of clear guidelines for ethical AI integration \cite{fuligni2025would, harvey2025don}. Additionally, high computational and infrastructure costs create accessibility barriers, limiting equitable access to AI-driven tools and widening learning gaps among students. Running large-scale AI models also raises energy consumption concerns, adding another layer of complexity to implementation \cite{belchev2025code}. Moreover, security, privacy, and data integrity risks emerge as critical issues, with the potential for LLMs to store or leak sensitive academic data and pose compliance challenges with educational data protection regulations \cite{yao2024survey}. The lack of institutional support for AI-driven curriculum transformation makes the transition even more challenging, as many educators lack proper training on how to effectively leverage AI without compromising learning objectives. In addition, the absence of regulatory frameworks for AI in education leaves many ethical and security questions unanswered, making institutions hesitant to adopt these technologies on a large scale \cite{schiff2022education}. To ensure sustainable and fair AI integration, universities must invest in AI literacy programs, provide equitable access to AI resources, and establish strict security measures that protect students' academic data while optimizing the benefits of AI-enhanced education.

    Overall, the integration of Large LLMs in software engineering education presents significant challenges across multiple dimensions, necessitating a balanced and well-regulated approach. Learning and educational challenges arise from LLMs’ limitations in contextual understanding, biases, hallucinations, and the risk of over-reliance, which may hinder students' critical thinking and independent problem-solving skills. Assessment and academic integrity concerns highlight the risks of AI-assisted plagiarism, ethical dilemmas in grading, and the potential for academic dishonesty, necessitating stronger detection mechanisms and redesigned assessment strategies. The impact on student skill development and cognitive load further complicates AI adoption, as excessive reliance on LLMs can diminish deep learning, critical reasoning, and problem-solving abilities, making it crucial to establish clear evaluation frameworks that ensure AI supports rather than replaces cognitive engagement. Integration and practical implementation challenges, including course redesign, high computational costs, and security risks, create further barriers, requiring institutions to invest in AI literacy programs, infrastructure, and data protection strategies. Collectively, these challenges underscore the need for a structured and ethical AI integration framework that ensures LLMs enhance rather than disrupt software engineering education, supporting students in developing real-world competencies while maintaining academic integrity and inclusivity.

\section{Implications and Validity Threats} \label{Sec:Implications}
We now summarise the key research and practical implications of this study results and the validity threats:

\subsection{Research Implications}
The identified motivators and demotivators of using LLMs in software engineering education provide critical insights into the evolving role of AI in academic settings. From a research perspective, the motivators, such as enhanced learning experiences, personalized education, and improved assessment mechanisms, highlight the potential of LLMs to revolutionize software engineering education. These findings contribute to existing literature by offering evidence on how AI-driven tools can support adaptive learning and improve student engagement. Additionally, the demotivators, including concerns about academic integrity, AI biases, over-reliance on automation, and security risks, present new challenges that require further exploration.

\subsection{Practical Implications}
The identified motivators suggest that institutions can leverage LLMs to enhance personalized learning, automate assessment, and support skill development in software engineering education. These insights provide a roadmap for integrating AI-driven tools into curricula while ensuring student engagement and pedagogical effectiveness. On the other hand, the demotivators highlight the need for proactive strategies to address ethical concerns, mitigate biases, and prevent over-reliance on AI-generated content. Universities and educators must establish clear policies on AI usage, promote academic integrity, and ensure that AI-assisted learning complements rather than replaces critical thinking and problem-solving skills. The findings also stress the importance of faculty training and AI literacy programs to equip educators with the necessary skills to implement and regulate AI-driven learning tools. 

Future research will be built on these findings to develop adaptive frameworks that will assist in bringing LLMs tools in Software Engineering education in Finnish higher education institutes. 

\subsection{Threats to Validity} \label{Sec:Threat_Validity)}
Various threats could affect the validity of the study findings. In this study, we analyzed the potential threats in four main categories—internal, external, construct, and conclusion validity—as defined by Wohlin \cite{wohlin2012experimentation}.

\subsubsection{Internal Validity}
Internal validity refers to whether the observed results are truly caused by the investigated factors rather than external influences. In this study, potential biases in data selection, researcher interpretations, and thematic mapping could pose threats. Additionally, using a traditional literature review instead of a systematic review (SLR) increases the risk of selection bias, incomplete coverage, and limited reproducibility. To mitigate these threats, we applied systematic data coding, cross-validated findings with external reviewers, documented the review process step by step, and assessed interpersonal bias using Kendall’s coefficient of concordance analysis.

\subsubsection{External Validity}
External validity refers to the extent to which the study findings can be generalized. In this study, the identified motivators and demotivators were derived from a traditional literature review, which may limit the breadth of included studies. Additionally, the thematic mapping, while structured, is influenced by selected studies and may not capture all possible themes. To mitigate this, we ensured diverse sources, followed a structured review process, and validated the thematic mapping through a formal thematic approach and multiple reviewers (authors) to reduce bias.

\subsubsection{Construct Validity} 
Construct validity assesses whether the study accurately measures the intended concepts. In this study, threats may arise from misclassification of motivators and demotivators or inconsistencies in thematic mapping, which could affect the reliability of identified themes. To mitigate this, we followed a formal thematic analysis process, used clearly defined coding schemes, and involved multiple reviewers to ensure consistency and reduce subjective bias.

\subsubsection{Conclusion Validity}
Conclusion validity ensures that the study’s findings and inferences are logically sound and well-supported by the data. Threats in this study may arise from biases in data interpretation, limited dataset coverage, or overfitting of thematic patterns, potentially leading to misleading conclusions. To mitigate these threats, we followed a structured review process, critically evaluated emerging patterns through consensus meetings among all authors, and ensured clear documentation of the review process, including a replication package \cite{khan2025motivators} for transparency and reproducibility.

\section{Related Work}
The integration of LLMs in education has been extensively studied, with various works focusing on their pedagogical applications, ethical concerns, and impact on student learning. However, while prior research provides valuable insights into the broader implications of LLMs in education, most studies do not specifically analyze the motivators and demotivators of LLM adoption in software engineering education within Finnish higher education institutions.

For instance, study conducted by Arto et al. \cite{hellas2023exploring} examines how LLMs such as OpenAI Codex and GPT-3.5 respond to programming help requests in an introductory course at Aalto University, Finland. This study evaluates the accuracy, reliability, and limitations of LLM-generated feedback, finding that while GPT-3.5 identifies errors effectively, it frequently produces false positives and unnecessary suggestions. The study highlights both the benefits and risks of LLM-assisted learning in programming education, but it does not investigate the broader challenges of integrating LLMs into structured curricula nor the pedagogical motivators and demotivators of their adoption in software engineering education.
Samuli et al. \cite{laato2023ai} explores the potential and challenges of AI-driven learning tools in Finnish university education. It systematically maps ChatGPT’s capabilities and identifies thirteen key implications for AI-assisted learning, including its benefits in personalized tutoring, accessibility, and interactive learning. However, the study also raises concerns about academic integrity, over-reliance on AI, and the need for educators to adapt teaching strategies to integrate LLMs effectively. While this work provides valuable insights into the general use of LLMs in higher education, it does not explicitly address their impact on software engineering education.
Reza \cite{fattahibavandpour2024advancing} conducts a systematic review of LLMs in education, focusing on their scalability, ethical concerns, and implementation strategies. This study examines the applications of AI chatbots, automated feedback, and personalized learning, while also addressing challenges related to bias, data privacy, and responsible AI use. The research presents a roadmap for integrating LLMs into Finland’s education system, particularly in the context of higher education institutions. However, it does not provide an examination of specific motivators and demotivators nor does it focus on software engineering students’ learning experiences and assessment-related challenges.
Melany et al. \cite{macias2024empowering} presents two LLM-based educational tools developed at Metropolia University of Applied Sciences in Finland. These tools include a Moodle\footnote{https://moodle.org/} AI plugin designed to assist teachers in managing course content and a curriculum planning tool that aligns course descriptions with strategic educational goals. This study takes a usability-focused, user-centric approach to integrating AI into existing educational systems while ensuring compliance with data protection regulations such as GDPR\footnote{https://gdpr-info.eu/}. While this research highlights the practical applications of AI-driven tools for educators, it does not investigate how LLMs impact students' learning experiences, particularly in software engineering education.

\subsection{Comparison with Proposed Work}
Now we provide the comparison between the proposed study and the related studies discussed in the previous section (See Table \ref{tab:RelatedWork}). The proposed study differentiates itself by reviewing the state-of-the-art literature on motivators and demotivators of using LLMs in software engineering education, providing a structured thematic mapping of factors that influence their adoption. Unlike previous studies that primarily focus on ChatGPT general applications in education or the usability of AI tools for educators, this research examines LLMs from the perspective of both students and educators in the domain of software engineering. It provides a literature review of 25 motivators and 30 demotivators, categorizing them into themes such as enhancing learning experiences, assessment and feedback, collaboration and peer learning, and skill development. Furthermore, the study extends beyond theoretical discussions by offering a roadmap (understanding of motivators and demotivator factors) that Finnish universities can use to integrate LLMs effectively into their software engineering curricula while addressing key concerns such as over-reliance, academic integrity, and ethical AI use.

\renewcommand{\arraystretch}{1.5}
\begin{table}
 \centering
\normalsize
\caption{Proposed Study Comparison With Related Studies. Note: (\checkmark: included, \xmark: not included, *: abstract overview, SE: Software Engineering)}
\label{tab:RelatedWork}
\resizebox{\columnwidth}{!}{%
 \begin{tabular}{|l|c|c|c|c|c|}
 \hline
 \textbf{Study} & \textbf{Focus Area} & \textbf{Motivators/Demotivators} & \textbf{SE Education} & \textbf{Finland Context} & \textbf{Review Paper} \\
\hline
\textit{Arto et al. \cite{hellas2023exploring} } & LLMs in coding education & \xmark & \cmark (*) & \cmark & \xmark \\
\hline
\textit{Samuli et al. \cite{laato2023ai}} & AI in university education & \xmark & \xmark & \cmark & \xmark \\
\hline
\textit{Reza \cite{fattahibavandpour2024advancing}} & AI Chatbots \& Business Models & \xmark & \xmark & \cmark & \cmark \\
\hline
\textit{Melany et al. \cite{macias2024empowering}} & AI tools for digital pedagogy & \xmark & \xmark & \cmark & \xmark \\
\hline
\textbf{Proposed Study} & \textbf{LLMs in SE Education} & \textbf{\cmark} & \textbf{\cmark} & \textbf{\cmark} & \textbf{\cmark} \\
\hline
\end{tabular}
}
\end{table}

\section{Conclusions and Future Work} \label{sec:Conlusions}
This study examined the integration of LLMs in software engineering education by identifying key motivators and demotivators from the literature. The findings reveal that LLMs provide several advantages, including personalized learning, enhanced engagement, and automated assessments, making education more efficient and adaptive. These benefits indicate that LLMs can significantly improve the learning experience by offering real-time feedback, supporting skill development, and facilitating collaborative learning environments. However, the study also uncovered major concerns, such as ethical issues, biases in AI-generated content, the risk of over-reliance on automation, and challenges in academic integrity. These demotivators highlight the need for a structured and well-regulated approach to integrating LLMs into education, ensuring that they complement rather than replace essential problem-solving and critical thinking skills.

Building on these findings, the next phase of this research will involve empirical studies, including surveys, interviews, and case studies, to validate and expand upon the insights gained from the literature review in the domain of Finnish higher education institutes. These studies will provide deeper insights into how LLMs are perceived and utilized in real-world educational settings, helping to refine the understanding of both their advantages and challenges. The ultimate objective is to develop a well-defined framework tailored for Finnish higher education institutions, offering guidelines for responsible and effective LLM integration. This framework will ensure that LLMs enhance teaching and learning outcomes while addressing potential risks and maintaining educational integrity. Through this structured approach, the research aims to contribute to the ongoing discussion on AI in education and support institutions in making informed decisions about LLM adoption.

 \section{Acknowledgments and Authors Contribution}
 The authors acknowledge the use of generative AI tools, such as ChatGPT, along with the writing assistant Grammarly, and writefull (Overleaf tool), to enhance the clarity and coherence of this paper. However, these tools were employed as aids rather than replacements for the authors input. All authors performed significant revisions and thorough reviews to refine the content, and the authors assume full responsibility for the final version of the manuscript.\\
 Maryam Khan (PhD Researcher) was primarily responsible for data collection, analysis, and reporting, with equal assistance from Muhammad Azeem Akbar (Co-Supervisor). The third author, Jussi Kasurinen (Main Supervisor), contributed to reviewing, writing, and finalizing the manuscript.
 
\bibliographystyle{ACM-Reference-Format}
\bibliography{Main}

\end{document}